\RequirePackage{amsmath}
\RequirePackage{amssymb}
\documentclass[runningheads]{llncs}

\usepackage[T1]{fontenc}
\usepackage{graphicx}
\usepackage{hyperref}

\begin{document}

\title{Analyzing Trendy Twitter Hashtags in the 2022 French Election}
\titlerunning{2022 French Election Trendy Hashtags Analysis}  % abbreviated title (for running head)
\author{Aamir Mandviwalla\inst{1,2} \and Lake Yin\inst{1,2} \and Boleslaw K. Szymanski\inst{1,2}}
\authorrunning{A. Mandviwalla, L. Yin, B.K. Szymanski} % abbreviated author list (for running head)
%\tocauthor{Aamir Mandviwalla, Lake Yin, and Boleslaw K. Szymanski}
%
\institute{Department of Computer Science, Rensselaer Polytechnic Institute,  Troy, NY 12180, USA \and Network Science and Technology Center, Rensselaer Polytechnic Institute, Troy, NY 12180, USA}

\maketitle              % typeset the title of the contribution
\let\thefootnote\relax\footnotetext{This preprint is a version of a publication of the same title presented at the {\it Complex Networks \& Their Applications XII Conference}, Nov. 28-30, 2023, Menton, France. Wording may vary between versions. The copyrighted final version is accessible at \url{https://doi.org/10.1007/978-3-031-53468-3_18}}
\begin{abstract}
Regressions trained to predict the future activity of social media users need rich features for accurate predictions. Many advanced models exist to generate such features; however, the time complexities of their computations are often prohibitive when they run on enormous data-sets. Some studies have shown that simple semantic network features can be rich enough to use for regressions without requiring complex computations. We propose a method for using semantic networks as user-level features for machine learning tasks. We conducted an experiment using a semantic network of 1037 Twitter hashtags from a corpus of 3.7 million tweets related to the 2022 French presidential election. A bipartite graph is formed where hashtags are nodes and weighted edges connect the hashtags reflecting the number of Twitter users that interacted with both hashtags. The graph is then transformed into a maximum-spanning tree with the most popular hashtag as its root node to construct a hierarchy amongst the hashtags. We then provide a vector feature for each user based on this tree. To validate the usefulness of our semantic feature we performed a regression experiment to predict the response rate of each user with six emotions like anger, enjoyment, or disgust. Our semantic feature performs well with the regression with most emotions having $R^2$ above 0.5. These results suggest that our semantic feature could be considered for use in further experiments predicting social media response on big data-sets.
\end{abstract}

\section{Introduction}
In recent years, social media data has been increasingly used to predict real-world outcomes. Data from platforms like Twitter, Reddit, and Facebook has been shown to be valuable in predicting public sentiment or response towards many different topics. This information has been used across many different fields like predicting stock market price changes or movie popularity \cite{5616710,DBLP:journals/corr/PagoluCPM16}.

Social media platforms have continued to get more popular over time. Due to this, the size of social media datasets continues to increase. As the size of these datasets gets bigger and bigger, computational time complexity of the algorithms being used becomes a significant issue. Some of the most popular features used in social media predictions like sentiment analysis become prohibitively expensive when working with larger datasets \cite{kumari2016impact,electronics11213624}.

In this paper we propose a method to generate features that can be used in social media predictions on big datasets. We create a weighted semantic network between Twitter hashtags from a corpus of 3.7 million tweets related to the 2022 French presidential election. A bipartite graph is formed where hashtags are nodes and weighted edges connect the hashtags reflecting the number of Twitter users that interacted with both hashtags. The graph is then transformed into a maximum-spanning tree with the most popular hashtag designated as its root node to construct a hierarchy amongst the hashtags. We then provide a vector feature for each user where the columns represent each of the 1037 hashtags in the filtered dataset and the value for each column is the normalized count of interactions for the user with that hashtag and any children of the hashtag in the tree.

To validate the usefulness of our semantic feature we performed a regression experiment to predict the response rate of each user with six emotions like anger, enjoyment, or disgust. The emotion data was manually annotated by a DARPA team created for the INCAS Program. We provide a baseline simple feature representing the counts the number of times a user interacts with each of the 1037 hashtags. Both the baseline and our semantic feature perform well with the regression with most emotions having $R^2$ above 0.5. The semantic feature outperforms the baseline feature on five out of six emotions with a $p$ value of $0.05$.

The rest of the paper is organized as follows. Section \ref{section:related} details related works. In Section \ref{section:data}, we present the dataset used for experimentation. Then, Section \ref{section:methods} describes the methodology used in our paper. The design of experiments and their results are presented in Section \ref{section:results}, and the conclusions are discussed in Section \ref{section:conclusion}.

\section{Related Works}\label{section:related}

Analyzing Twitter using semantic networks has been done in the past with various methods to determine relationships between hashtags and their trends. For example, \cite{shi} considered two hashtags to be semantically related if an individual tweet contained both hashtags in the text. Similarly, \cite{featherstone} created a semantic network based on word co-occurrence within tweets. However,~\cite{radicioni} presented an approach using a bipartite network between users and hashtags where an edge between a user node and a hashtag node was added if the user tweeted the hashtag at least once. This bipartite network was then projected into a monopartite network of hashtags. This approach is more applicable to our purposes because it captures the latent social network of the dataset. In addition, \cite{radicioni} focuses on a Twitter dataset taken from the 2018 Italian elections which is similar to our 2022 French election dataset.

Many studies have shown that semantic network features can be rich enough to use for regressions in a multitude of situations. In the field of psychology, semantic networks can be used to analyze a person’s vocabulary to gain insight on cognitive states \cite{chan,beckage}. In terms of social media semantic networks, \cite{grassi} used semantic networks generated from sentences as features for a time series regression to capture the volatility of the stock market. In general, these approaches involve creating a semantic network for each person or object in the study. The alternative approach is to create large-scale, singular semantic networks that can be used to describe all users. For example, \cite{he} demonstrated a recommender system which used a large-scale word co-occurrence semantic network created from social media posts to recommend related social media posts to users. Such approach might be better for analyzing users since it can take advantage of the nuanced relationships between different social media communities, which cannot be done with an approach that only generates an individual semantic network for each user.

\section{Data}\label{section:data}

We applied these enrichments to a dataset provided by DARPA INCAS program team that comprised 3.7 million French language tweets from 2022. This dataset was collected such that each tweet is relevant to the discussions that arose during the 2022 French presidential election. After pruning, this dataset contains 1037 hashtags and 389,187 users.

\section{Methods}\label{section:methods}

\subsection{Semantic Network Generation}

We performed several steps to prepare the Twitter data and create a semantic network.

\subsubsection{Preprocessing}\label{sec:method1}

The corpus of Tweets was first cleaned by removing URLs with regular expression and French stop words using the NLTK Python library \cite{nltk}. Each Tweet was tokenized by converting all words to lowercase, removing digit-only words, and removing punctuation, except for hashtags. After extracting a set of hashtags and corresponding occurrence counts, any hashtags with an occurrence count below the mean were removed from the set to focus on trendy hashtags.

\subsubsection{Bipartite Graph Generation}

 Using this set of hashtags, a bipartite graph was constructed between users and the hashtags where an edge indicates an interaction between a user and a hashtag. Following the technique introduced in \cite{radicioni}, we implemented the bipartite graph as an adjacency list where a set of interacted hashtags is stored with each user. A user is said to \emph{interact} with a trendy hashtag if the user retweets, quotes retweets, comments under, or posts a tweet that contains the hashtag word with or without the hashtag symbol. We chose this relaxed approach because we consider situations such as "france" versus "\#france" to be semantically identical. The resulting bipartite graph was projected along the hashtags as a weighted semantic network where each node represents a trendy Twitter hashtag, and the weighted edges represent the shared audience of users between two trendy hashtags. 
 
 \subsubsection{Edge Pruning}

\begin{figure}
\includegraphics[width=\textwidth]{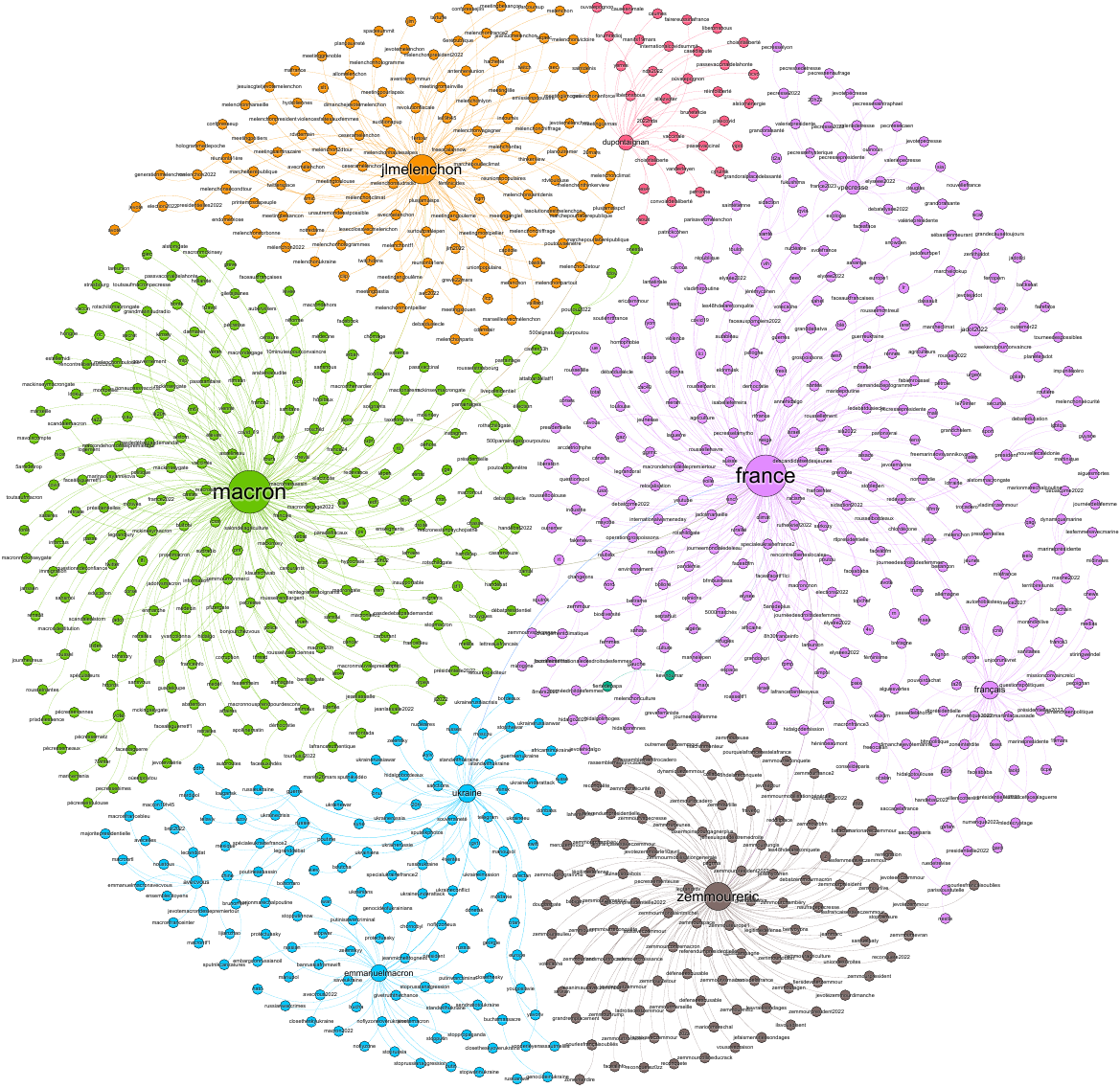}
\caption{\textbf{Maximum-Spanning Tree of trendy hashtags in the 2022 French Presidential Election}. Nodes represent trendy hashtags. Weighted edges between nodes represent the number of users that interacted with both trendy hashtags. Node size is based on weighted degree. Node color is based on modularity class after applying Louvain community detection and provided to help distinguish groups of similar nodes visually. From the root node \#france, a hierarchy amongst trendy hashtags is formed, with clear distinction between different presidential candidates.}
\label{fig:1}
\end{figure}

 Next, the bipartite graph was then converted into a maximum spanning tree (MST) to only consider the most important links between trendy hashtags. We had conducted multiple experiments with and without edge pruning and concluded that some form of edge pruning is essential for removing noisy edges. We tested a flat cutoff approach for excluding edges with a weight below a set cutoff, and the MST approach, achieving the best and most robust results with the MST. All graph operations were performed with the NetworkX Python library \cite{networkx}. A visualization of the resulting MST can be seen in Figure \ref{fig:1}.

\subsection{Semantic User Enrichment}\label{sec:method2}

Each user in the dataset $u \in U$ is assigned a set containing the trendy hashtags they had interacted with using the previously described interaction criteria. For each user set $S_u$, and for each trendy hashtag $t \in S_u$ where $S_u \subset V$ given semantic network/graph $G(V, E)$, each adjacency list corresponding to $t$  is converted into an adjacency vector $\textbf{a}_t=(a_{t1},\dots,a_{tn})$ where $n = |V|$, and

\[a_{ti} = \begin{cases}
    \frac{w(t, m(i))}{c(t)} & \text{if } e(t, m(i)) \in E \\
    1              & \text{if } m(i) = t \\
    0                 & \text{otherwise}
\end{cases}\]

where $w:E \to \mathbb{N}$ is the weight of the edge $e(u, v) \in E$, $m:[0,n] \to V$ that maps each index to a trendy hashtag, and $c:V \to \mathbb{N}$ maps each trendy hashtag to the number of users that have interacted with it. The set of vectors for each user is then summed element-wise and then normalized by dividing by the $L^2$ norm of the summed vector. The result for each user is a vector representing this user interests in a trendy hashtag and related trendy hashtags weighted by the latent social network. 

\subsection{Baseline User Enrichment}\label{sec:method3}

To judge the utility of our semantic network enrichment, we devised a simpler baseline enrichment for comparison. Each user $u \in U$ is assigned a vector $\textbf{a}_u=(a_{u1},\dots,a_{un})$ where $n = |V|$, and $a_{ui} = k(u, m(i))$ where 
\[k(u, t) = \begin{cases}
    1 & \text{if $u$ interacts with trendy hashtag $t$} \\
    0 & \text{otherwise}
\end{cases}\]
Each vector is normalized by dividing by the sum of the vector elements.

\section{Results}\label{section:results}

\subsection{Regression Experiment}
To compare the enrichments, we conducted an experiment to test the performance of the enrichments in a regression task. We decided to test if a user enrichment could be used to predict the average "emotions" for each user. Each tweet in the dataset was annotated by the DARPA team with an array of six distinct emotions and an "other" value (representing fear, anger, enjoyment, sadness, disgust, surprise, and "none of the above" tag) where the sum of each array equals 1. For every user's tweets, we summed the emotion arrays, then divided the resulting array by its 1-norm, so that each array element follows $U(0,1)$ and represents the probability of that user interacting with each emotion. Each user array was split into a set of emotion target variables, and each one was paired with the corresponding user enrichment method as the input variable. Only users with $\geq 10$ tweets were included, resulting in 49,360 entries of input/target pairs for each emotion. Since this experiment is only meant to compare the different methods relative to each other with often minimal differences, we used the Scikit-learn implementation of linear regression \cite{scikit}.

\subsection{Experiment Results}

\begin{table}[h]
\centering
\begin{tabular}{|c|c|c|}\hline

Emotion & {Baseline $R^2$} & {Semantic $R^2$} \\
\hline

Fear & 0.222 & 0.229 \\
\hline

Anger* & 0.567 & 0.574 \\
\hline

Enjoyment* & 0.634 & 0.648 \\
\hline

Sadness* & 0.266 & 0.277 \\
\hline

Disgust* & 0.501 & 0.514 \\
\hline

Surprise* & 0.082 & 0.098 \\
\hline

None & 0.416 & 0.423 \\
\hline

\end{tabular}
\caption{\label{tab:2} Regression experiment results. (*: semantic > baseline; $p < 0.05$)}
\end{table}

Overall, there is a clear pattern of improved performance when using the semantic enrichment instead of the baseline as seen in Table \ref{tab:2}. All specific emotions, except fear and "none of the above" tag, statistically significantly improved performance using the semantic method versus the baseline. Statistical significance was calculated using an F-test between the two models. 

Interestingly, the baseline linear regression performed poorly on surprise emotion. The semantic enrichment performed significantly better, but still was weakly correlated with response. Intuitively, people would have various positive or negative views towards certain political trendy hashtags, which would correlate with most of the emotions. However, surprise cannot easily be categorized as on the positive or negative binary spectrum, which could explain why the linear regression performed poorly in those cases. Previous research on sentiment analysis has also shown notably lower performance when predicting surprise \cite{buechel2016emotion,tsakalidis2018building}.

\subsection{Semantic Analysis}

To analyze which trendy hashtags are most associated with improvement with the semantic enrichment, we decided to filter for the top 10\% of users that saw the most improvement in prediction accuracy between the baseline regression and the semantic regression. This was determined by the mean absolute error in emotion predictions versus the ground truth. Then we compared trendy hashtag occurrence rates for the top 10\% users with the hashtag occurrence rates for the rest of the users. All hashtag occurrence rates were calculated based on direct interactions, like the baseline enrichment. We then selected the top 10 trendy hashtags that saw the largest increase in occurrence rate between the top 10\% of users and the rest of the users. 

\begin{table}[h]
\centering
\begin{tabular}{|c|c|c|}\hline

Trendy hashtag & Bottom 90\% rate & Top 10\% rate \\
\hline

Paris & 0.274 & 0.420 \\
\hline

Youtube & 0.328 & 0.470 \\
\hline

Europe & 0.458 & 0.600 \\
\hline

Lci & 0.227 & 0.348 \\
\hline

Passe & 0.250 & 0.370 \\
\hline

Ukrainians & 0.287 & 0.404 \\
\hline

Jeunes & 0.282 & 0.397 \\
\hline

Liberté & 0.371 & 0.485 \\
\hline

Nucléaire & 0.284 & 0.394 \\
\hline

Immigration & 0.248 & 0.357 \\
\hline

\end{tabular}
\caption{\label{tab:3} Trendy hashtags with the largest increase in occurrence rate between the top 10\% of users and the rest of users. Occurrence rate is the proportion of users that directly interacted with that trendy hashtag. The top 10\% of users is computed based on the improvement in prediction accuracy between the baseline regression and the semantic regression for those users.}
\end{table}

Since the presence of these trendy hashtags in Table \ref{tab:3} result in more accurate emotion predictions with the semantic enrichment, this suggests that the users engaging with these trendy hashtags tend to engage with other emotionally salient trendy hashtags, which would be more useful when predicting emotion levels. Given the severity of war, it would make sense that "Ukrainians" would be strongly connected to other highly emotionally salient trendy hashtags. A previous English language Twitter study about the Ukrainian war found that "Ukrainians" is a significant buzzword, so it is not surprising that the word reappears in French. In addition, that study identified the YouTube twitter account that was frequently mentioned in relation to the war, which would explain why it invokes strongly emotional trendy hashtags \cite{a16020069}.

\section{Conclusions and Future Work}\label{section:conclusion}
We have connected semantic networks to the area of machine learning, demonstrating using a simple experiment that this can be used to consistently improve results on real-world data. To the best of our knowledge, this is the first time a semantic network feature for machine learning has been explored.

Future work can include specializing in such a framework to tackle specific problems. It is worth noting that the semantic network method used in this paper was designed with the constraints of the INCAS challenge in mind, which might not necessarily be the best way to utilize semantic networks when describing users in other situations. We are reporting these results simply to show that this method is a notable improvement over simpler approaches and it is worth investigating other applications in future work. One of the main drawbacks of this approach is the increased computational time associated with projecting the bipartite graph into a monopartite semantic network. However, there are distributed computing approaches to monopartite projection challenges, so this can be scaled for large scale applications involving many trendy hashtags \cite{8566259}.

During testing, we found that pruning small edges is important for removing noise from the final enrichments. We used the most aggressive pruning approach, by using a maximum spanning tree to only retain the strongest edges. This has the disadvantage of removing connections between different communities within the graph. It is possible that using a more sophisticated pruning approach could improve the quality of using semantic networks in this manner. Future work can take inspiration from knowledge graph edge pruning methods, which can account for domain information of the trendy hashtags \cite{10.5555/3304222.3304334}.

\subsubsection{Acknowledgements} {This work was partially supported by the DARPA INCAS Program under Agreement No. HR001121C0165 and by the NSF Grant No.
BSE-2214216}


\bibliographystyle{splncs04}

\bibliography{main}

\end{document}